\documentclass[a4paper, twocolomns, 10pt]{article}
\usepackage[left=19mm, right=19mm, bottom=33mm, top=33mm]{geometry}
\pdfoutput=1
\pdftrue

\usepackage{amsmath}
\usepackage[pdftex]{graphicx}

\usepackage[T2A]{fontenc}
\usepackage[utf8]{inputenc}
\usepackage[english]{babel}
\usepackage{multicol}
\usepackage{colortbl} 

\usepackage{geometry}
\usepackage{amsfonts}
\usepackage{amsmath}
\usepackage{amssymb}
\usepackage{ntheorem}
\usepackage{latexsym}
\usepackage{units}
\usepackage{textcomp}
\usepackage{wrapfig}
\usepackage{booktabs}
\usepackage{hyperref}

\usepackage{xcolor}
\hypersetup{
    colorlinks,
    linkcolor={red!50!black},
    citecolor={blue!50!black},
    urlcolor={blue!80!black}
}

\everymath{\displaystyle}
\usepackage{color}
\everymath{\displaystyle}

\renewcommand{\phi}{\varphi}

\usepackage{url}

\usepackage{sectsty} 
\sectionfont{\fontsize{11}{14}\selectfont}     
\subsectionfont{\fontsize{10}{13}\selectfont}  

\begin{document}
\begin{flushleft}
\footnotesize  
\rmfamily      
\begin{tabular}[c]{l}
\rowcolor[gray]{.9}This is a preprint version of an article published in Eur.\hspace{0.5mm}Phys.\hspace{0.5mm}J.\hspace{0.5mm}C\hspace{0.4mm}\hspace{0.4mm}84,\hspace{0.2mm}266(2024).\\
\rowcolor[gray]{.9}The final authenticated version is available online at: \href{https://doi.org/10.1140/epjc/s10052-024-12444-y}{https://doi.org/10.1140/epjc/s10052-024-12444-y}
\end{tabular}
\end{flushleft}
\vspace{20mm}

\begin{multicols}{2}
\end{multicols}

\vspace{-30mm}
\section*{\Large Searches for the light invisible axion-like particle in \boldmath{$K^{+}$$\to\pi^{+}\pi^{0}a$} ~decay}
\begin{center}
\begin{minipage}{1.0\linewidth}
  {\center \large \textsc{The OKA collaboration}\\}\vspace{-2mm}
\end{minipage}
\end{center}
\begin{center}
\begin{minipage}{1.0\linewidth}
 \center{
  \textsc
  A.~S.~Sadovsky{${}^{a}$},
  A.~P.~Filin,
  S.~A.~Akimenko{$^{\dagger}$},
  A.~V.~Artamonov,
  A.~M.~Blik{$^{\dagger}$},\\
  S.~V.~Donskov,
  A.~M.~Gorin,
  A.~V.~Inyakin,
  G.~V.~Khaustov,
  S.~A.~Kholodenko,\\
  V.~N.~Kolosov,
  V.~F.~Kurshetsov,
  V.~A.~Lishin,
  M.~V.~Medynsky,
  V.~F.~Obraztsov,\\
  A.~V.~Okhotnikov,
  V.~A.~Polyakov,
  V.~I.~Romanovsky,
  V.~I.~Rykalin,
  V.~D.~Samoylenko,\\ 
  I.~S.~Tiurin,
  V.~A.~Uvarov,
  O.~P.~Yushchenko
 }\vspace{-3mm}
 \center{\small 
   \textsc{(NRC "Kurchatov Institute"${}^{}_{}{}^{}$-${}^{}_{}{}^{}$IHEP, 142281 Protvino, Russia),} 
 }\vspace{-1mm}
 \center{
  \textsc
  S.~N.~Filippov,
  E.~N.~Gushchin,
  A.~A.~Khudyakov,
  V.~I.~Kravtsov,\\
  Yu.~G.~Kudenko{${}^{b,c}$},
  A.~V.~Kulik,
  A.~Yu.~Polyarush,
 }\vspace{-3mm}
 \center{\small 
   \textsc{(Institute for Nuclear Research -- Russian Academy of Sciences, 117312 Moscow, Russia),} 
 }\vspace{-1mm}
 \center{
  \textsc
  V.~N.~Bychkov, 
  G.~D.~Kekelidze,
  V.~M.~Lysan,
  B.~Zh.~Zalikhanov
 }\vspace{-3mm}
 \center{\small 
   \textsc{(Joint Institute of Nuclear Research, 141980 Dubna, Russia)}\\
 }\vspace{-1mm}
\end{minipage}
\end{center}

{
\footnotesize
\line(1,0){170}\\
\vspace{-1mm}${}$\hspace{0.8cm}${}^{a}$~e-mail: Alexander.Sadovskiy@ihep.ru\\
\vspace{-1mm}${}$\hspace{0.8cm}${}^{b}$~Also at National Research Nuclear University (MEPhI), Moscow, Russia\\
\vspace{-1mm}${}$\hspace{0.8cm}${}^{c}$~Also at Institute of Physics and Technology, Moscow, Russia\\
\vspace{-1mm}${}$\hspace{0.8cm}${}^{\dagger}$~Deceased
}
\vspace{-4mm}
\begin{center}
\begin{minipage}{0.09\linewidth}
~
\end{minipage} 
\begin{center}
\begin{minipage}{0.83\linewidth}
{ 
  \rmfamily
  {\bf Abstract}
  A high-statistics data sample of the $K^{+}$ decays is recorded by the OKA collaboration.
  A missing mass analysis is performed to search for a light invisible pseudoscalar axion-like particle (ALP)
  $a$ in the decay $K^{+} \to \pi^{+} \pi^{0} a$.
  No signal is observed, and the upper limits for the branching ratio of the decay are calculated. 
  The $90\%$ confidence level upper limit changes from 
  $2.5\cdot10^{-6}$ to $2\cdot10^{-7}$ for the ALP mass from 0 to 200 MeV/$c^{2}$, except for the region of $\pi^{0}$ mass,
  where the upper limit is $4.4\cdot10^{-6}$.
}\vspace{2mm}
{\\  {\bf Keywords} {Kaon decays~$\cdot$~exotic particles~$\cdot$~axion~$\cdot$~axion-like particles~$\cdot$~experimental results}}
\end{minipage}
\end{center}
\begin{minipage}{0.09\linewidth}
~
\end{minipage}
\end{center}
\vspace{-2mm}

\begin{multicols}{2}


\section{Introduction}\label{SectInitro}
\vspace{-3.0pt}
The QCD Axion is a hypothetical pseudoscalar particle, that was postulated in \cite{PecceiQuinn} to solve the strong CP problem.
Its properties are described by the decay constant $f_{a}$, related to the scale of the Peccei-Quinn (PQ) symmetry breaking scale $\Lambda_{PQ}$:
$f_{a} = \Lambda_{PQ}/4\pi$.
The QCD axion mass is $m_{a} \sim m_{\pi}f_{\pi}/f_{a}$.
It decays to two photons, the decay time is $\tau_{a}$$\sim$$2^{8}\pi^{3}f^{2}_{a}/(\alpha m^{3}_{a})$.
When the axion is considered as a candidate dark matter particle, its lifetime $\tau_{a}$ is expected to be comparable to or greater than the lifetime of the Universe $\sim$ 13.8~Gyr, thus its mass estimate is $m_{a} \lesssim$ 10~eV/$c^{2}$.
Axion also has couplings to quark currents, in particular to the {\it sd} flavour-changing neutral current (FCNC). 
There is a vector and axial vector coupling: $\mathcal{L} = q_{\mu} a \{\bar{d}  (\gamma_{\mu}/F^{V}_{sd} + \gamma_{\mu}\gamma_{5}/F^{A}_{sd}) s \}$, 
where $q_{\mu}$~\textendash~axion four-momentum, $F^{A}_{sd}$ and $F^{V}_{sd}$ effective constants, with $F^{A/V} = 2f_{a}/C^{A/V}$ being model-dependent constants.

Due to parity conservation in QCD the decay  $K^{+} \to \pi^{+} \pi^{0} a$ is sensitive to the axial-vector coupling while much better constrained decay $K^{+} \to \pi^{+} a$ \cite{CortinaGil_NA61} tests the vector coupling.
More general models of axion-like particles (ALPs) consider cases in which the axion mass is not set by QCD dynamics only, but by some other mechanism.
These models have two free parameters: $m_{a}$ and $f_{a}$. Then there is a softer limit on axion mass: $m_{ALP} < 1$~GeV/$c^{2}$ \cite{P_Lo_ChiattoUNIMORE}.

The only result on the search for an axion in the $K^{+} \to \pi^{+} \pi^{0} a$ decay mentioned in PDG \cite{PDG} is that of the BNL E787 experiment \cite{BNL787}.
Better upper limits can be extracted from ISTRA+ paper \cite{oTchikilevISTRA2004}, devoted to the search for a pseudoscalar sgoldstino. The result is $\sim$ $Br<10^{-5}$ at $90\%$ C.L.

This  article is dedicated to a new search for the decay of $K^{+} \to \pi^{+} \pi^{0} a$. We assume that the ALP has a sufficiently long lifetime and decays outside the detector.
In our study, we rely on \cite{Camalich_PRD102, P_Lo_ChiattoUNIMORE}, where the phenomenology of the $K^{+} \to \pi^{+} \pi^{0} a$~ decay is considered in detail.

\section{The OKA setup}
\vspace{-5.0pt}
The {\bf OKA} is a fixed target experiment dedicated to the study of kaon decays using the decay in flight technique. 
It is located at NRC ''Kurchatov Institute''-$^{}$IHEP in Protvino (Russia). 
A secondary kaon-enriched hadron beam is obtained by RF separation with the Panofsky scheme.
The beam is optimized for a momentum of 17.7~GeV/c with a kaon content of about 12.5\% and an intensity up to $5\times 10^{5}$ kaons per U-70 spill.
\begin{figure*}[!ht]
\begin{center}
\includegraphics[width=1.00\textwidth]{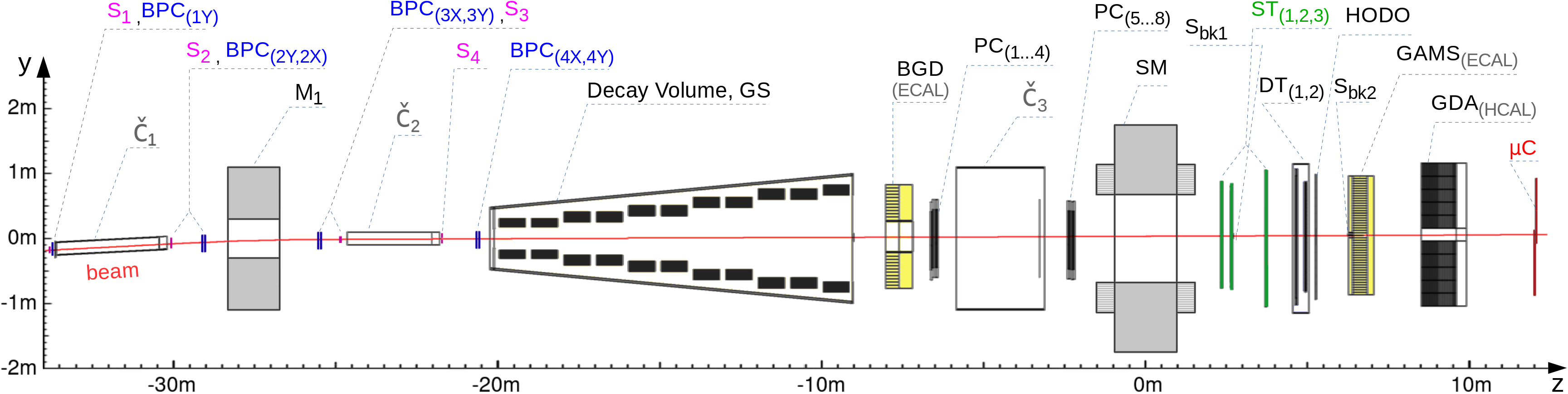}
\caption{\label{FigOkaSetup}Schematic elevation view of the OKA setup. See the text for details.}
\end{center}
\end{figure*}

The OKA setup (Fig.~\ref{FigOkaSetup}) makes use of two magnetic spectrometers on either side of an 11~m long {\small Decay Volume (DV)} filled with helium at atmospheric pressure
and equipped with a guard system ({\small GS}) of lead-scintillator sandwiches mounted in 11 rings inside the {\small DV} for photon veto.
It is complemented by an electromagnetic calorimeter {\small BGD} \cite{BGDref1982} (with a wide central opening). 

The first magnetic spectrometer measures the momentum of the beam particles (with a resolution of $\sigma_{p}/p$ $\sim 0.8\%$).
It consists of a vertically (y) deflecting magnet {\small M$_{1}$} surrounded by a set of 1~mm pitch (beam) proportional chambers {\small BPC${}_{(}$${}_{1Y,}$\hspace{-1pt} ${}_{2Y,}$\hspace{-1pt} ${}_{2X,...,}$\hspace{-1pt} ${}_{4Y}$${}_{)}$}. 
To measure the charged tracks from decay products, the second magnetic spectrometer is used (with resolution of $\sigma_{p}/p \sim$ 1.3\textendash2\% for momentum range of 2\textendash14 GeV/$c$). 
It consists of a wide aperture 200$\times$140 cm$^2$ horizontally (x) deflecting magnet {\small SM} with {\small $\int$}$Bdl \sim 1$~Tm surrounded by tracking stations:
proportional chambers {\small PC$_{(1,...,8)}$}, straw tubes {\small ST$_{(1,2,3)}$}, and drift tubes {\small DT$_{(1,2)}$}.
A matrix hodoscope {\small HODO} is used to improve time resolution and to link $x$\textendash$y$ projections of a track.

At the end of the setup there are: 
an electromagnetic calorimeter {\small GAMS$_{\tt(ECAL)}$} of 18X${}_{0}$ (consisting of $\sim$ 2300 3.8$\times$3.8$\times$45~cm$^3$ lead glass blocks) \cite{GAMSref1985},
a hadron calorimeter {\small GDA$_{\tt(HCAL)}$} of 5$\lambda$ (constructed from 120 20$\times$20~cm${}^{2}$ iron-scintillator sandwiches with WLS plates readout)
and a wall of four $1\times1$m${}^{2}$ muon counters {\small {$\mu$}C} mounted behind the hadron calorimeter. 

More details on the OKA setup can be found in \cite{OKA_status_2009, aFilinOkaDAQ, OKA_KmuHnu_EPJC}.

\section{The data and the analysis procedure}
\vspace{-4.0pt}
Two sequential sets of data with a beam momentum of 17.7\hspace{-0.5pt} GeV/c recorded by the OKA collaboration in 2012 and 2013 are analyzed to search for the ALP.

The present analysis is based on a trigger that selects kaon decays: ${\tt Tr_{Kdecay}=}$ ${\tt S_{1}{\cdot}S_{2}{\cdot}S_{3}{\cdot}S_{4}{\cdot}\overline{\check{C}}_{1}{\cdot}\check{C}_{2}{\cdot}\overline{S}_{bk} }$.
Four scintillation counters ({\small {S}$_{1}$, {S}$_{2}$, {S}$_{3}$, and {S}$_{4}$}) are used to select beam particles.
A combination of two threshold Cherenkov counters ({\small \v{C}$_{1}$} sees pions, {\small \v{C}$_{2}$} sees pions and kaons) is needed to select kaons.
An anti-coincidence with two scintillation counters ({\small S$_{bk1}$ or S$_{bk2}$}) located on the beam axis behind the {\small SM} magnet is used to suppress the recording of undecayed beam particles.
The trigger additionally requires an energy deposition in the {\small GAMS} e.m.~calorimeter higher than ~2.5 GeV
to suppress the dominating $K^{+}\to\mu^{+}\nu$ decay:
${\tt Tr_{GAMS}=Tr_{Kdecay} \cdot (E_{GAMS}>2.5}$~${\tt GeV})$.

Monte Carlo (MC) simulations of particle interactions with the detector and its response are performed using a software package based on the Geant-3.21 library \cite{Geant321}, which includes a realistic description of the setup.
The response of the trigger used in the experiment is also included in the simulation.
The MC events are passed through the full OKA reconstruction procedure.
Monte Carlo events for various decays use weights proportional to the square of the absolute value of the corresponding matrix element.
Signal events of $K^{+}\to\pi^{+}\pi^{0} a$ decay for ten masses of the axion from 0 to 210 MeV/c${}^{2}$ are generated. The matrix element is taken from \cite{Camalich_PRD102} and \cite{P_Lo_ChiattoUNIMORE}.

To estimate the background, samples of Monte Carlo events are used for the five main decay channels of charged kaon with $\pi^{0}$ in final state
($\pi^{+}\pi^{0}$, $\pi^{+}\pi^{0}\gamma$, $\pi^{+}\pi^{0}\pi^{0}$, $\pi^{0}\mu^{+}\nu$, and $\pi^{0}e^{+}\nu$) mixed according to their branching fractions.
The generated MC statistics are $\sim$ 8 times larger than the data sample recorded in the experiment.
The weights for the 3-body decays are calculated according to PDG \cite{PDG}. 
Processes in which the beam kaon scatters or interacts while passing through the setup are also added.

\begin{figure*}[!ht]
\begin{center}
\includegraphics[width=1.00\textwidth]{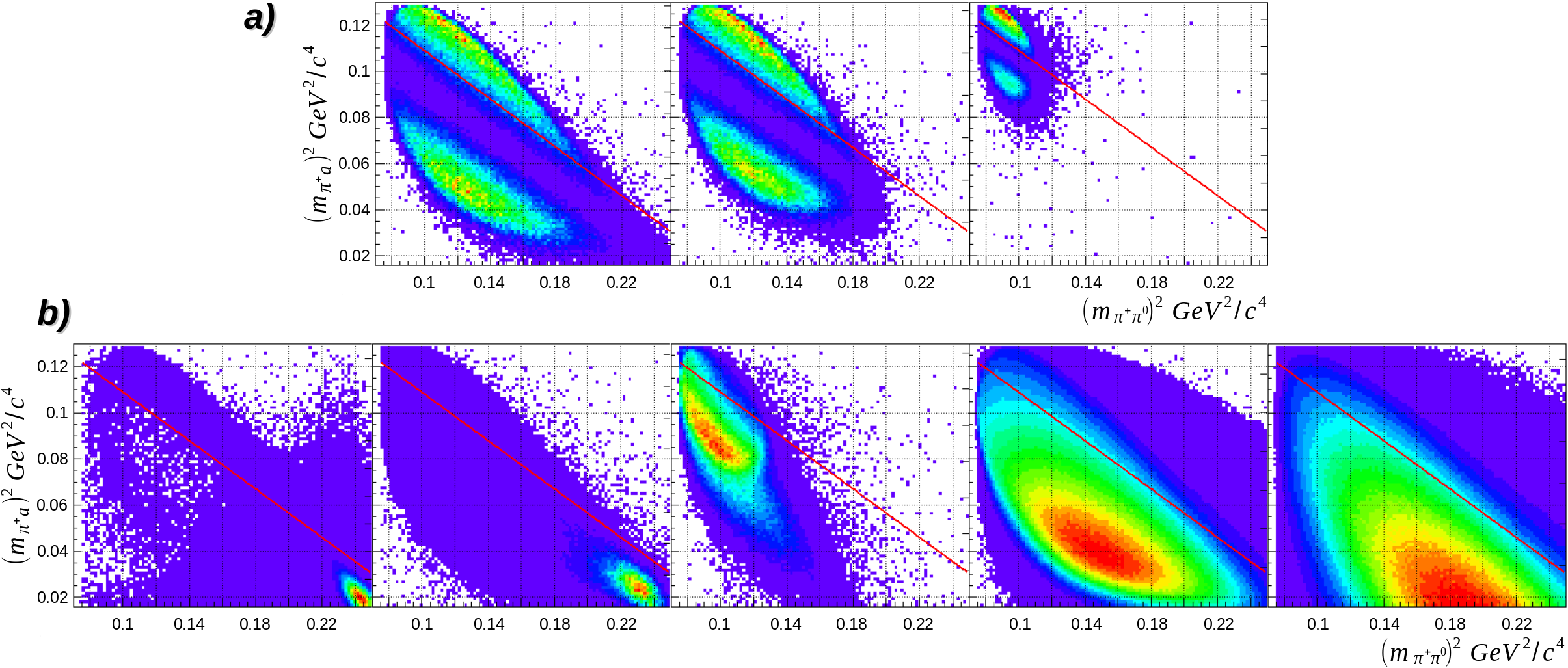}
\vspace{-8mm}\caption{\small
    The distribution of {\small  $(m_{\pi^{+}{a}})^{2}{^{~}}${vs.}${^{~}}(m_{\pi^{+}\pi^{0}})^{2}$ } for reconstructed events of the axion (plots {$\pmb a$}{\bf \it )}~) for a set of masses $m_{a} = \{0, 60, 160\}$~MeV/c${}^{2}$,
    and (plots {$\pmb b$}{\bf \it )}~) for the main background processes ($K^{+}\to\pi^{+}\pi^{0}$, $K^{+}\to\pi^{+}\pi^{0}\gamma$, $K^{+}\to\pi^{+}\pi^{0}\pi^{0}$, $K^{+}\to\mu^{+}\nu_{\mu}\pi^{0}$, and $K^{+}\to{e}^{+}\nu_{e}\pi^{0}$)
    according to MC simulation with the matrix elements.
    The selection of the area above the red line is applied to suppress the background processes strongly while keeping about half the signal due to the feature of its matrix element.
}
\label{SignalAndBg_Dalitz}
\end{center}
\end{figure*}

\subsection{Event selection}
\vspace{-3.0pt}
The total number of $\sim$ $3.65\times 10^{9}$ events with kaon decays is logged, of which $\sim$ $8\times 10^{8}$ events are reconstructed with a single charged particle in the final state.

The event selection for the decay process $K^{+}\to\pi^{+}\pi^{0}{a}$ starts with the requirement of a single beam track and a single secondary track with a decay angle $>4$ mrad and with a vertex matching distance (CDA) below $1.25$~cm.
A moderate chi-square cut for the charged track quality is applied.
To clean the tracking sample, it is required that no extra track segments behind the {\small SM} magnet are found.
The vertex position is required to be inside the {\small DV}.
The beam particle momentum is required within a range of $17.0<p_{beam}<18.6$~GeV/c. 
To select $\pi^{0}$ among decay products, the number of showers in electromagnetic calorimeters ({\small GAMS} or {\small BGD}) not associated with the track must be equal to 2.
For these events, the $\pi^{0}$ identification is done by invariant mass selection: $|m_{\gamma\gamma}-m_{\pi^{0}}|<15$~MeV/c$^{2}$.

After those selections, we obtain $30.3\cdot 10^{6}$ $K^{+}\to\pi^{+}\pi^{0}$ events in two runs ($18.4\cdot 10^{6}$ for 2012 and $11.9\cdot 10^{6}$ for 2013).
At this stage, the normalization of MC statistics to the data is performed.

\begin{figure*}[!ht]
\begin{center}
\includegraphics[width=1.01\linewidth]{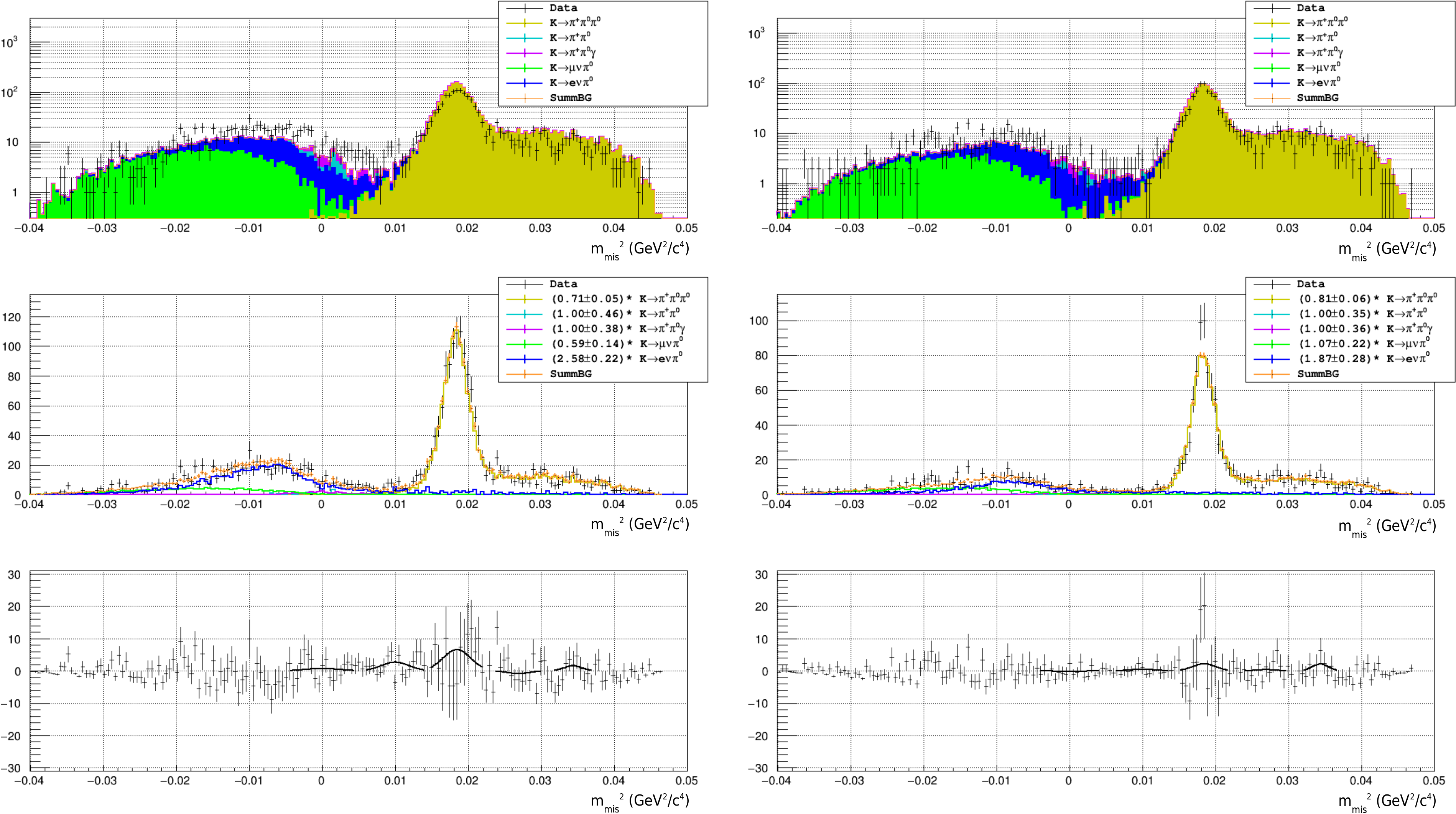}
\vspace{-8mm}\caption{
   \small
   {\tt{Upper plots}}: the resulting $m^{2}_{mis}$-distribution for the 2012 run (left) and 2013 run (right). The data is shown as black points with errors.
   Different background channels are marked with colors; the stack plot is used to highlight strongly suppressed processes (note the {\it log-y} scale).
   The normalization of the MC background is done using the number of $K^{+}\to\pi^{+}\pi^{0}$ events before the main cuts, and the sum of background channels is indicated with orange.
   {\tt{Middle plots}}: the result of the tuning of the relative magnitudes for the background processes. The corresponding scaling factors are depicted with their fit errors.
   The {\tt{bottom plots}} demonstrate the difference between the experiment and the sum of the tuned background processes.
   For illustration, the results of the signal fit are shown for five positions of $m^{2}_{mis}$.
}
\label{mm2distrib_r14_PsgAxion}
\end{center}
\end{figure*}

\subsection{Selecting the decay of interest}\label{MainCutsSelectingAxion}
\vspace{-3.0pt}
The Dalitz plots in the $(m_{\pi^{+}\pi^{0}})^2, (m_{\pi^{+} a})^2$ plane for the selected MC events (for three masses for the signal and for five main background decays) are shown in Fig.~\ref{SignalAndBg_Dalitz}.
The figures correspond to the first run, and the corresponding figures for the second run look very similar.

To allow for better separation of signal from background, the area below the red line on the Dalitz plot is rejected.
Strong suppression of the background is achieved, while the signal is reduced by a factor of $\sim$ 2 only. 

In order to disentangle $K^{+}\to\pi^{+}\pi^{0} a$ signal from its main backgrounds $K^{+}\to\pi^{+}\pi^{0}$ and $K^{+}\to\pi^{+}\pi^{0}\pi^{0}$, three kinematic cuts are applied:\\
the missing energy is $E_{mis}$$=$$(E_{K^{+}} - E_{\pi^{+}} - E_{\pi^{0}})$$>$$2.8$~GeV; 
and the momenta of $\pi^{0}$ and $\pi^{+}$ in the kaon rest frame are: $p^{*}_{\pi^{0}}<150$~MeV/c and $p^{*}_{\pi^{+}}<189$~MeV/c.

To suppress misidentified events from $K^{+}\to\mu^{+}\nu_{\mu}\pi^{0}$, a requirement of absence of signal from the muon counter $\mu\tt{C}$ matched with the charged track is used.

A further suppression of $K^{+}\to{e}^{+}\nu_{e}\pi^{0}$ and $K^{+}\to\mu^{+}\nu_{\mu}\pi^{0}$ is done with the help of calorimetry.

To suppress $e^{+}$, the events with $E/p>0.83$ ($E$ is the energy of the {\small GAMS} shower, associated with the track of momentum $p$) are discarded. 
Then, if the number of cells in the shower is $N>4$ or the cluster energy is $E>1.9$~GeV, the track is identified as $\pi^{+}$ with an "early" hadron shower in {\small GAMS}. 
Otherwise, one shower in {\small GDA}, matching the track within 22~cm and with either the number of cells in {\small GDA}, $N_{GDA}>4$, or with a sufficiently high ratio $E_{GDA}/p>0.67$ for which the $N_{GDA}>1$ is required. 
This cut rejects muons and selects pion with a "late" shower in {\small GDA}.

Finally, the total energy deposition in {\small GS} below the noise threshold of 100~MeV is required to suppress events with photons escaping the acceptance of electromagnetic calorimeters.

\subsection{Fits of the missing mass spectrum}
\vspace{-6.0pt}
As the next step, a search for the signal in the missing mass squared spectrum in the range 0~\textendash~0.05~GeV${}^{2}$/${c^{4}}$ is performed, $m^2_{a} = m^2_{mis} = ({\bf p}_{K^{+}} - {\bf p}_{\pi^{+}} - {\bf p}_{\pi^{0}})^{2}$, 
where ${\bf p}_{K^{+}}$, ${\bf p}_{\pi^{+}}$ and ${\bf p}_{\pi^{0}}$ are four-momenta of the corresponding particles.

The resulting $m^{2}_{mis}$-distributions for the data and for the main background processes are shown in Fig.~\ref{mm2distrib_r14_PsgAxion}. 
Due to strong suppression cuts ($\sim 10^{5}$), the efficiency estimates for background processes are known with noticeable errors, so a (maximum likelihood) fit
in which the efficiencies of the main background processes are allowed to vary is used to tune the background model to the experimental data. 
As $K^{+}\to\pi^{+}\pi^{0}\pi^{0}$ background is indistinguishable at this stage        from the signal with $m=m_{\pi^{0}}$ the $2\sigma$ region around $(m_{\pi^{0}})^{2}$ is excluded
from the fit, and we are using only combinatorial background tails as an estimate for this background magnitude.
The variations          for $K^{+} \to \pi^{+} \pi^{0}$ and $K^{+} \to \pi^{+} \pi^{0} \gamma$ backgrounds during the fit procedure are only allowed to have lower values to prevent these processes from substituting the potential signal in the low-mass region.

As a result of the fit we obtain a reasonable description of the background in the wide $m^{2}_{mis}$-mass range, see middle and bottom plots in Fig.~\ref{mm2distrib_r14_PsgAxion}, 
the residual discrepancy mainly affects the negative non-physical side of the $m^{2}_{mis}$-distribution, 
but due to the non-ideal resolution for the signal at zero mass this affects the extraction of the signal near $m^{2}_{mis} \approx 0$.
\begin{wrapfigure}[8]{l}{1.0\linewidth} 
\vspace{-3ex}
\includegraphics[width=1.00\linewidth]{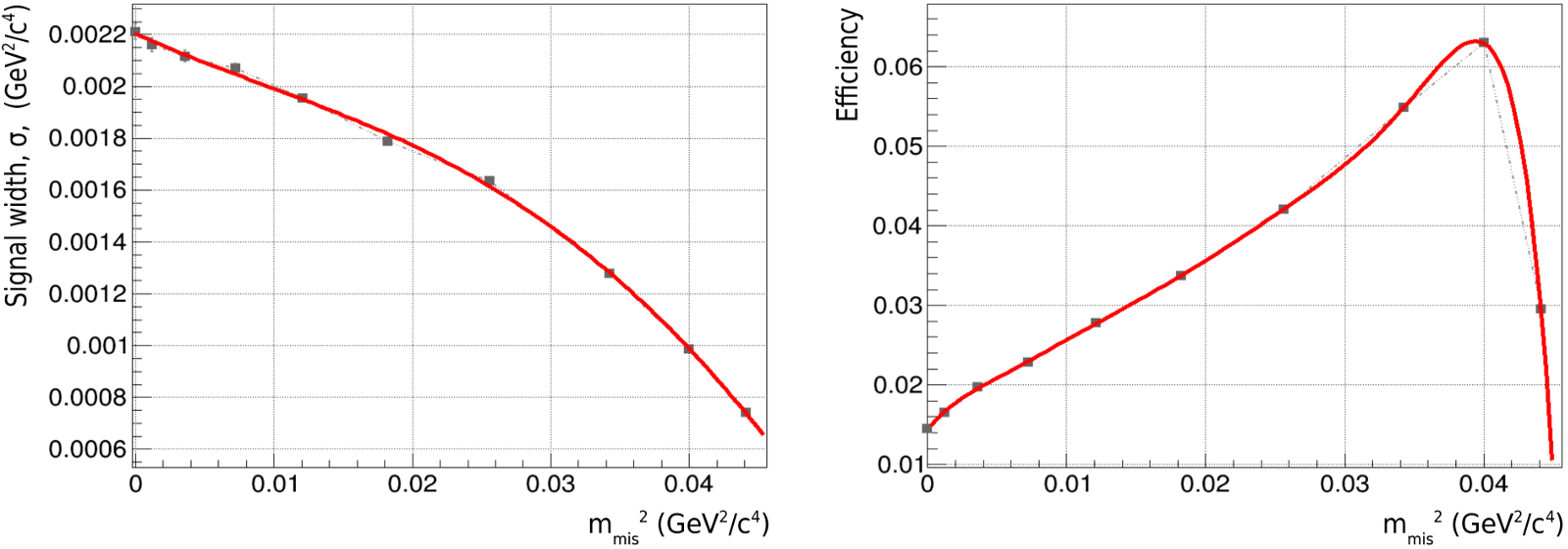}
\vspace{-8mm}
\caption{\small
 The width ($\sigma$ of the $m^{2}_{mis}$ distribution) and efficiency of the signal are calculated for 10 mass positions (black points) 
and shown together with their interpolation curves (shown in red).
 These figures correspond to the first run; the width of the signal for the second run is $\sim 10\%$ less, while the efficiency is almost the same. 
}
\label{FigSignalWidthEfficiency}
\end{wrapfigure}
~\\ ~\\ ~\\ ~\\ ~\\ ~\\ ~\\ 
\vspace{20mm}

The next stage is a maximum likelihood (ML) fit of the distributions shown in Fig.~\ref{mm2distrib_r14_PsgAxion} (middle plots) with a tuned background in which signal magnitude is extracted separately at 89 different positions of $m^{2}_{a}$.

\begin{figure*}[!ht]
\begin{center}
\includegraphics[width=1.00\textwidth]{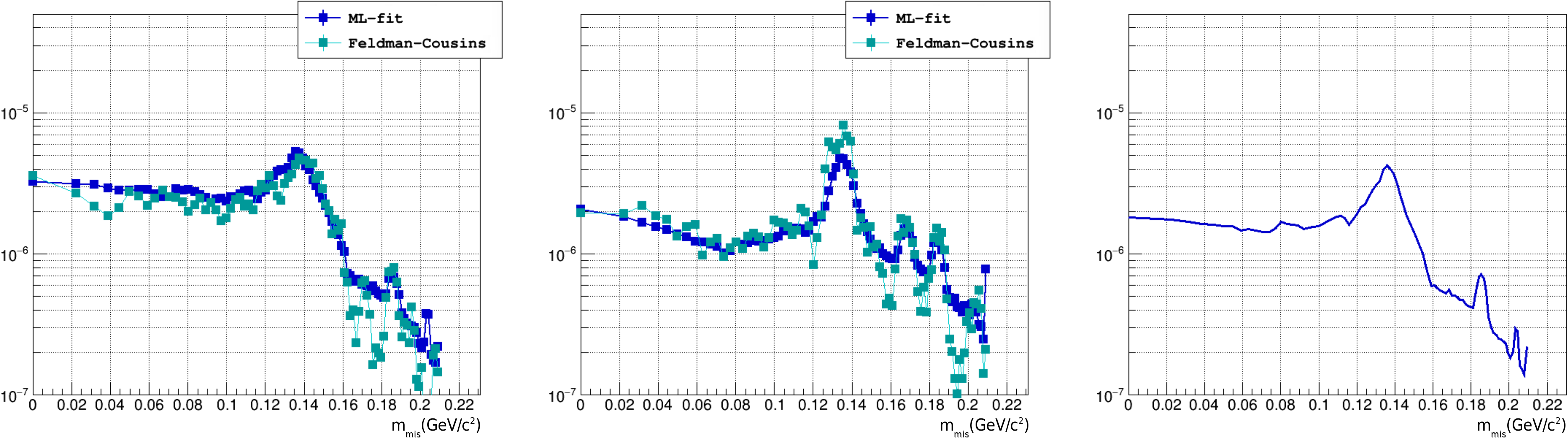}
\vspace{-8mm}
\caption{\small
    The 90{\%} CL upper limits on $K^{+}\to\pi^{+}\pi^{0}{a}$ branching (left \textendash ~for 2012 run, middle \textendash ~for 2013 run, right \textendash ~for the statistical sum of the runs).
    The Feldman-Cousins method \cite{FeldmanCousinsMethod} is used for the comparison, see text for details.
}
\label{Result_r14r15Sum_ULatCL90_PsgAxion}
\end{center}
\end{figure*}

According to the MC simulation, the signal $m^{2}_{mis}$ distribution is well described by the Gaussian over a wide range of axion masses.
The $\pm 2\sigma$ range is used to fit the signal at each $m^{2}_{a}$ point, with the width $\sigma$ taken from the MC; it decreases from $\sim$ $2\cdot10^{-3}$ to $\sim 1\cdot10^{-3}$ GeV$^{2}/c^{4}$ within the considered mass range, see Fig.~\ref{FigSignalWidthEfficiency} (left). 
Complementary to that, Fig.~\ref{FigSignalWidthEfficiency} (right) shows the efficiency\footnote{The efficiencies given in the article are normalized to the number of kaon decays in the nominal length of the DV.} dependence.

In the absence of a statistically significant signal in the considered mass region, we estimate the upper limit for the number of signal events.
We follow the method used in \cite{oTchikilevISTRA2004}, in which the fit procedure for the signal is not bound to positive values only (to avoid a possible bias).
The variation of the signal width by 5\% is used during the fit procedure to account for possible systematics in the MC signal width.
The one-sided upper limit for the number of signal events corresponding to the confidence level (CL) of 90\% is constructed as $N_{UL@90{\%}CL} = max(N_{P},0)+1.28\cdot\sigma_{N_{P}}$,
where $N_{P}$ and $\sigma_{N_{P}}$ are evaluated from the ML fit, being the number of signal events and its error.
This approach is chosen because  it allows one to profit from the knowledge of the signal shape, in contrast to the Feldman-Cousins \cite{FeldmanCousinsMethod} method, where only the central region of the signal is used. 
The additional advantage is that the obtained parameters $N_{P}$ and $\sigma_{N_{P}}$ are convenient for the statistical combination of two runs.
Nevertheless, for comparison, we present the results with the Feldman-Cousins method (where the window of $\pm 1.2 \sigma$ for the signal is applied); see Fig.~\ref{Result_r14r15Sum_ULatCL90_PsgAxion} (left, middle).
The statistical sum of the two experiments is shown in Fig.~\ref{Result_r14r15Sum_ULatCL90_PsgAxion} (right).

\subsection{Branching calculations, systematic errors study}
\vspace{-3.0pt}
The upper limits for the branching ratio are calculated relative to the well-measured $K^{+}\to\pi^{+}\pi^{0}$ (aka $K\pi2$) decay \cite{PDG}. 
It is expected that the main sources of systematics related to the trigger, track quality, and particle identification would cancel out in this approach.
It is  also  true for the {\small GS} (veto) cut, as we apply it to $K\pi2$ events at this stage.

\begin{figure*}[!ht]
\begin{center}
\includegraphics[width=1.00\textwidth]{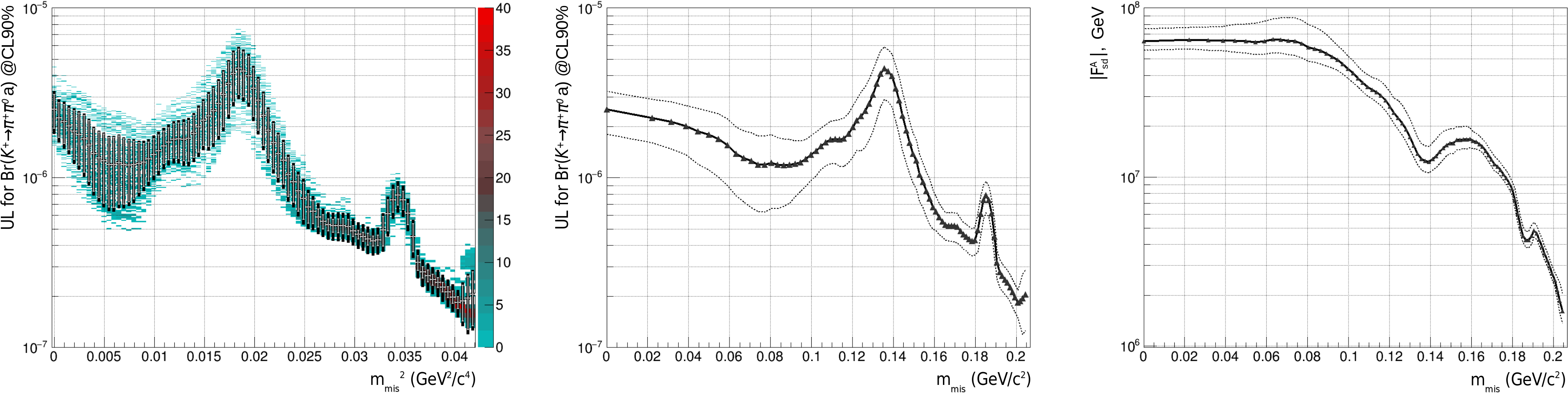}
\vspace{-6mm}
\caption{\small
    The 90{\%} CL upper limits for the $K^{+}\to\pi^{+}\pi^{0}{a}$ branching (left and middle figures).
    The scatter plot (left figure) demonstrates the systematics arising from the variation of the cuts used for the selection (flat distribution of 20 variables within $\pm 1\sigma$). 
    The RMS of the distribution at each $m^{2}_{mis}$ is indicated by gray bars superimposed over the scatter plot, while
    the remaining systematic errors are added quadratically and shown with black bars.
    The middle plot demonstrates the final result for the ULs. The black triangles correspond to the mean value obtained from the variation of the selection criteria, and the dotted lines correspond to the systematic errors.
    The right plot indicates the corresponding lower limit on the $|F^{A}_{sd}|$ parameter, together with the corresponding systematic error.
}
\label{FinalResult_ULatCL90_PsgAxion}
\end{center}
\end{figure*}

The branching for the signal decay is calculated as follows:\\
$Br(K^{+}$$\to$$\pi^{+}\pi^{0}{a})=Br(K\pi2)\cdot (N_{a}/\varepsilon_{a})\cdot(\varepsilon_{K\pi2}/N_{K\pi2})$, 
where the efficiency  and number of events for the run 2012 (2013) are:
$\varepsilon_{K\pi2}=6.9\cdot10^{-2}$ ($7.2\cdot10^{-2}$) and $N_{K\pi2}=9.7\cdot10^6$ ($6.1\cdot10^6$) events;
$\varepsilon_{a}$ is the signal efficiency, taken from the parametrization shown in Fig.~\ref{FigSignalWidthEfficiency} (right).
The systematic error in $\varepsilon_{a}$ is about 5\%.
The systematic error of the $(\varepsilon_{K\pi2}/N_{K\pi2})$ ratio is estimated as $3.5\%$, while the statistical one is negligible.

A systematic error of {14.8\%} is estimated, taking into account the difference between the $K^{+}\to\pi^{+}\pi^{0}$ and $K^{+}\to\pi^{+}\pi^{0}\pi^{0}$ (with lost $\pi^{0}$, which is detected by the missing mass spectrum) normalization alternatives.
For the $K^{+} \to \pi^{+}\pi^{0}\pi^{0}$ case, the veto ({\small GS}) cut is removed to avoid suppression of the second (escaping) $\pi^{0}$, and also (in contrast to the $K^{+}\to\pi^{+}\pi^{0}$), the $E_{mis}$ cut is used.

A systematic error arising from the theory due to errors in experimentally measured $K_{e4}$ formfactors \cite{Batley_NA48_2_PLB2012}, used in the signal MC, is $\lesssim$ 5\%.
It is estimated via reanalysis with the signal MC, in which two formfactors with the most significant errors, $f_{p}$ and ${g'}_{p}$, were varied accordingly in different combinations.

To account for possible systematic errors related to selection criteria, we repeated over a hundred standard analyses described earlier in this paper,
where a set of main cuts was randomly chosen within a window of $\pm 1\cdot \sigma_{j}$ around the original values ($\sigma_{j}$ is the resolution of a $j$-cut variable, estimated from the MC).
The resulting combinations of UL~vs.~$m^{2}_{mis}$ are shown in Fig.~\ref{FinalResult_ULatCL90_PsgAxion} (left) as a scatter plot.
The mean upper limit at each mass point is considered as the final result, and the obtained RMS of the distribution is treated as the systematic error of the upper limit, shown with gray bars superimposed in Fig.~\ref{FinalResult_ULatCL90_PsgAxion} (left).

Finally, systematic errors on the:        $(\varepsilon_{K\pi2}/N_{K\pi2})$,             $\varepsilon_{a}$,         normalization, and                                             $K_{e4}$ formfactors are added quadratically to the RMS obtained from the variation of the selection criteria 
and the result is shown by the black bars in Fig.~\ref{FinalResult_ULatCL90_PsgAxion} (left).

The final upper limit with the systematic errors included is shown in Fig.~\ref{FinalResult_ULatCL90_PsgAxion} (middle). 
The corresponding lower limit for the effective constant $|F^{A}_{sd}|$ is calculated accordingly and shown in Fig.~\ref{FinalResult_ULatCL90_PsgAxion} (right).

\vspace{-3mm}
\section*{Conclusions}
\vspace{-5.0pt}
The OKA data is analyzed to search for the light axion-like particle.
A peak search method in the missing mass spectrum is used in the analysis. 
No signal is observed, and the upper limits for the branching ratio in the mass range 0\textendash200~MeV/c$^{2}$ are set.
The current best upper limits for the branching ratio, Br($K^{+}\to\pi^{+}\pi^{0}{a}$) can be obtained from the paper of ISTRA+ \cite{oTchikilevISTRA2004}, which was originally devoted to the search for the pseudoscalar sgoldstino.
The only result on the axion search in $K^{+}\to\pi^{+}\pi^{0}{a}$  decay mentioned in PDG \cite{PDG} is that of the BNL-787 experiment \cite{BNL787}.
It should be noted, that both searches were performed under the uniform phase-space distribution hypothesis,
which appeared to be a rough approximation, as can be seen from Fig.~\ref{SignalAndBg_Dalitz}. 
Our analysis uses the realistic matrix element from \cite{Camalich_PRD102, P_Lo_ChiattoUNIMORE} and improves the limit \cite{oTchikilevISTRA2004} by a factor of 3.5\textendash10, depending on the axion mass.
Using the expression from \cite{P_Lo_ChiattoUNIMORE}, relating $F^{A}_{sd}$ and the branching, we calculated the lower limits for $|F^{A}_{sd}|$, shown in  Fig.~\ref{FinalResult_ULatCL90_PsgAxion} (right). 
The limit is about $6.4 \cdot 10^7$~GeV for the axion mass below 70~MeV/c${}^{2}$, which is, according to \cite{Camalich_PRD102}, the best limit for $|F^{A}_{sd}|$ among the HEP experiments.

\small
\subsection*{Acknowledgements}
\vspace{-3.0pt}
We express our gratitude to our colleagues in the accelerator department for the good performance of the U-70 during data taking; 
to colleagues from the beam department for the stable operation of the 21K beamline, including RF-deflectors, and to colleagues 
from the engineering physics department for the operation of the cryogenic system of the RF deflectors.
We are grateful to Vladimir Anikeev for fruitful discussions related to systematic errors.\\
This work was supported by the RSCF grant {\it{N\textsuperscript{\underline{\scriptsize o}}}22-12-0051}.

\end{multicols}
\end{document}